\documentstyle[epsf,eqsecnum,floats,aps]{revtex}
\begin{document}
\title{Antisymmetrised $2p$-forms generalising
curvature $2$-forms and a corresponding
$p$-hierarchy of Schwarzschild type metrics in dimensions $d>2p+1$
\footnote{CPHT 729.0799}}

\author{{\large A. Chakrabarti}\footnote{e-mail: chakra@cpht.polytechnique.fr}}
\address{{\small Centre de Physique Th\'eorique\footnote{Laboratoire Propre du
CNRS UMR7644},
Ecole Polytechnique, F-91128 Palaiseau, France}}
\author{ {\large D. H. Tchrakian}\footnote{e-mail: tigran@thphys.may.ie}}
\address{{\small Department of Mathematical Physics,
 National University of Ireland Maynooth, Maynooth, Ireland}}
\address{{\small and}}
\address{{\small School of Theoretical Physics -- DIAS, 10 Burlington 
Road, Dublin 4, Ireland }}

\maketitle
\medskip
\medskip

\date{}
\newcommand{\dd}{\mbox{d}}\newcommand{\tr}{\mbox{tr}}
\newcommand{\ee}{\end{equation}}
\newcommand{\be}{\begin{equation}}
\newcommand{\ii}{\mbox{i}}\newcommand{\e}{\mbox{e}}
\newcommand{\pa}{\partial}\newcommand{\Om}{\Omega}
\newcommand{\vep}{\varepsilon}
\newcommand{\bfph}{{\bf \phi}}
\newcommand{\lm}{\lambda}
\renewcommand{\thefootnote}{\fnsymbol{footnote}}
\newcommand{\re}[1]{(\ref{#1})}
\newcommand{\bfR}{{\sf R\hspace*{-0.9ex}\rule{0.15ex}%
{1.5ex}\hspace*{0.9ex}}}
\newcommand{\N}{{\sf N\hspace*{-1.0ex}\rule{0.15ex}%
{1.3ex}\hspace*{1.0ex}}}
\newcommand{\Q}{{\sf Q\hspace*{-1.1ex}\rule{0.15ex}%
{1.5ex}\hspace*{1.1ex}}}
\newcommand{\C}{{\sf C\hspace*{-0.9ex}\rule{0.15ex}%
{1.3ex}\hspace*{0.9ex}}}
\renewcommand{\thefootnote}{\arabic{footnote}}

\begin{abstract}
Starting with the curvature $2$-form a recursive construction of totally
antisymmetrised $2p$-forms is introduced, to which we refer as $p$-Riemann
tensors. Contraction of indices permits a corresponding
generalisation of the Ricci tensor. Static, spherically symmetric ``$p$-Ricci
flat'' Schwarzschild like metrics are constructed in this context for
$d>2p+1$, $d$ being the spacetime dimension. The existence of de Sitter type
solutions is pointed out. Our $2p$-forms vanish for $d<2p$ and the limiting cases
$d=2p$ and $d=2p+1$ exhibit special features which are discussed briefly. It is
shown that for $d=4p$ our class of solutions correspond to double-selfdual Riemann
$2p$-form (or $p$-Riemann tensor). Topological aspects of such generalised
gravitational instantons and those of associated (through spin connections)
generalised Yang-Mills instantons are briefly mentioned. The possibility of a study
of surface deformations at the horizons of our class of ``$p$-black holes'' leading
to Virasoro algebras with a $p$-dependent hierarchy of central charges is commented
on. Remarks in conclusion indicate directions for further study and situate our
formalism in a broader context.
\end{abstract}
\medskip
\medskip
\newpage

\section{Introduction}

The basic aim of this work is to present Schwarzschild like solutions to a
hierarchy of gravitational systems that we studied sometime ago~\cite{CT,OT},
which generalise the Einstein-Hilbert system in a very natural way. In this framework
the Lagrangian is higher order in the Riemann tensor, but in such a way that only
the quadratic power of the velocity fields (namely derivatives of the metric or the
{\it vielbein}) appears in it. This formalism opens up broader possibilities of
constructing gravitational instantons in suitable higher dimensions. Here we present
another remarkable aspect, which is one more evidence of the aptitude of our generalisation.
The spherically symmetric, static black hole (or the Schwarzschild metric) which is {\it unique}
in the Einstein-Hilbert framework, is the first member of a hierarchy in our context.
De Sitter type soluions also exist, rather systematically.

In the following sections we will deal with totally antisymmetrised $2p$-forms
constructed from the Riemann tensor, to which we refer as Riemann $2p$-forms or
$p$-Riemann tensors. The explicit recursive prescription is given below. Throughout
we will consider spacetimes of arbitrary dimension $d$, with $(d-1)$ spatial
dimensions. Our $2p$-forms vanish for $d<2p$ and the dynamics leads to
very special, degenerate cases for $d=2p$ and $d=2p+1$. The generic situation arises
for $d>2p+1$, $(p=1,2,3,...)$.

Contracting indices of the Riemann $2p$-forms one obtains the standard Ricci tensor
for $p=1$ and a generalised hierarchy of ``$p$-Ricci'' tensors to be defined below.
We show that for spherical symmetry one obtains as
``$p$-Ricci flat'' solutions a remarkable hierarchy of ``$p$-black holes'',
generalising the Schwarzschild metric in $d$ dimensions as follows
\[
\dd s^2 = -N_{(p)} \dd t^2 +N_{(p)}^{-1} \dd r^2 +r^2 \dd \Omega_{(d-2)}^2 \: ,
\]
with
\be
\label{1}
N_{(p)}=1-\frac{2C}{r^{\frac{d-2p-1}{p}}} \: ,
\ee

\noindent
$\dd \Omega_{(d-2)}^2$ being the line element on the unit $(d-2)$-sphere. For $p=1$
one recovers the standard lapse function in $d$ dimensions~\cite{MP},
\be
\label{2}
N_{(1)}\equiv N=1-\frac{2C}{r^{d-3}} \: .
\ee

The derivation of the solution \re{1} and the study of various aspects of interest
will be given below.

Let us present the construction of the $p$-Ricci tensors. A detailed account can be
found in \cite{OT} where many sources are cited. We present here the essential steps
using throughout differential forms in the frame-vector ({\it vielbein}) basis which
will be most useful for our purposes.

In the following sections we will be concerned with a particular, simple, class of
metrics. Generally, for some metric, a suitable set of tangent frame vectors are
the $1$-forms
\[
e^a =e_{\mu}^a \: \dd x^{\mu} \: ,
\]
where as usual $a,b,...$ denote frame indices and $\mu , \nu , ...$ space-time ones.

The torsionless antisymmetric spin-connection $1$-forms
\[
\omega^{ab} =\omega_{\mu}^{ab} \dd x^{\mu} =-\omega^{ba}
\]
satisfy
\be
\label{3}
\dd e^a +\omega^{ab} \wedge e_b =0 .
\ee

The curvature tensor is now given by the antisymmetric $2$-form
\be
\label{4}
R^{ab}=\dd \omega^{ab}+\omega^a{}_c \wedge \omega^{cb} =-R^{ba} \: ,
\ee
which can be expressed in the ``$e$-basis'' as
\be
\label{5}
R^{ab}=R_{a' b'}^{ab} \: e^{a'} \wedge e^{b'} .
\ee

\re{5} is the Riemann $2$-form corresponding to $p=1$. Now we introduce the following
recursive construction of totally antisymmetric Riemann $2p$-forms.

For $p=2$ we define
\be
\label{6}
R^{abcd} =R^{ab}\wedge R^{cd}+R^{ad}\wedge R^{bc}+R^{ac}\wedge R^{db} \: ,
\ee
having cyclically permuted $(b,c,d)$. Using the antisymmetry of $R^{ab}$ and, the
fact that $R^{ab} \wedge R^{cd} =R^{cd} \wedge R^{ab}$ since $2$-forms commute, one
readily verifies that $R^{abcd}$ is totally antisymmetric.

For $p=3$ we define
\be
\label{7}
R^{abcdef}=R^{ab}\wedge R^{cdef}+R^{af}\wedge R^{bcde}+R^{ae}\wedge R^{fbcd}
+R^{ad}\wedge R^{efbc}+R^{ac}\wedge R^{defb}
\ee
having cyclically permuted $(b,c,d,e,f)$. There are 15 terms of the type
$R^{ab}\wedge R^{cd}\wedge R^{ef}$ in \re{7} and again it can be checked readily
that $R^{abcdef}$ is totally antisymmetric.

The general recursion scheme is
\be
\label{8}
R^{a_1 a_2 ...a_{2p}}=
R^{a_1 a_2}\wedge R^{a_3 a_4 ...a_{2p}} +
{\rm cyclic\:  permutations \: of} \: (a_2 ,a_3 ,...,a_{2p})\: ,
\ee
which consists of $3.5...(2p-3)(2p-1)$ terms of the type $R^{a_1 a_2}\wedge
R^{a_3 a_4}\wedge ...\wedge R^{a_{2p-1} a_{2p}}$, and is totally antisymmetric. In the
$e$-basis
\be
\label{9}
R^{a_1 a_2 ...a_{2p}}=R_{b_1 b_2 ...b_{2p}}^{a_1 a_2 ...a_{2p}}\:
e^{b_1}\wedge e^{b_2}...\wedge e^{b_{2p}}.
\ee

For $p=1$ the Ricci tensor is given by
\be
\label{10}
{R_{(1)}}_b^a =\sum_c R_{bc}^{ac} =R_b^a \: .
\ee
We define the $p$-Ricci tensor to be
\be
\label{11}
{R_{(p)}}_{b_1}^{a_1}={1\over (2p-1)!}\sum_{(a_2 ,...,a_{2p})}
R_{b_1 a_2 ...a_{2p}}^{a_1 a_2 ...a_{2p} } \: ,
\ee
such that the $p$-Ricci scalar pertaining to it is
\be
\label{10a}
R_{(p)}=\sum_{a}{R_{(p)}}_a^a \: .
\ee
The overall constant $(2p-1)!$ in \re{11} takes acount of the number of permutaions
of the indices summed over
subject to the total antisymmetry of both upper and lower indices. This normalisation
factor, though convenient, is however not essential for our purposes.

The corresponding generalisations of the Einstein-Hilbert Lagrangian and the Einstein
tensor are given in \cite{OT}. But, apart from pointing out in passing how de Sitter
type solutions arise very simply and evidently, our concern here will be limited to
$p$-Ricci flat solutions.

In the absence of matter and cosmological constant the Lagrangian is
$\sqrt{|g|} R_{(p)}$ and the variational minima are obtained
(in the $e$-basis)~\cite{OT} from
\be
\label{12}
{R_{(p)}}_b^a -\frac{1}{2p}g_b^a R_{(p)}=0.
\ee
For the $p$-Ricci flat case, consistently with $R_{(p)}=0$, the solutions satisfy
\be
\label{12a}
{R_{(p)}}_b^a =0 \: .
\ee
One computes \re{11} for a given metric introduced as an ansatz and involving one (or
possibly more) unknown function(s) to be consistently determined to satisfy the
constraints \re{12a}.

In this paper we study the {\it spherically symmetric} solutions, in which case it is
natural to start (since we have the hierarchy \re{1} in mind) with
\be
\label{13}
\dd s^2 = -N \dd t^2 +N^{-1} \dd r^2 +r^2 \dd \Omega_{(d-2)}^2 \: ,
\ee
and see if one can obtain an $N$ satisfying all the constraints \re{12a}. In the event,
we will start with the Kerr-Schild (K-S) form of the metric given by
\begin{eqnarray}
g_{\mu \nu} &=& \eta_{\mu \nu} +2Cl_{\mu} l_{\nu}  \nonumber \\
g^{\mu \nu} &=& \eta^{\mu \nu} -2Cl^{\mu} l^{\nu} \label{14}
\end{eqnarray}
where $\eta_{00}=-1 , \eta_{ij}=\delta_{ij}$, (i=1,2,...,d-1), and
\be
\label{15}
l^{\mu}l_{\mu} =\eta^{\mu \nu}l_{\nu} l_{\mu}=g^{\mu \nu}l_{\nu} l_{\mu}=0 \: .
\ee
For black hole type solutions with a horizon, one assumes $C>0$.

For spherical symmetry one may set
\be
\label{16}
l_i =l_0 \frac{x_i}{r} =l_o \hat x_i \: ,
\ee
with $r=\sqrt{|x_i|^2}$ and $|\hat x_i|^2 =1$.

Now \re{13} and \re{14} are related through
\be
\label{17}
x_0 =t+\int \frac{\dd r}{N} -r \: ,\qquad L\equiv l_0^2 ={1\over 2C}(1-N) \: .
\ee

The K-S form is non-diagonal and thus loses some simple properties of \re{13}. But we
consider its introduction worthwhile for the following two reasons.

\begin{itemize}

\item
In \re{13} there are {\it two} distinguished coordinates $(t,r)$ while in \re{14}
there is only {\it one}, $x_0$, all the space-coordinates being treated on the same
footing. As will be shown below, this leads to a smaller number of groups of terms to
be summed over. (In fact, for spherical coordinates each angular one, $\theta_i$, also
contributes differently through $\sin \theta_i$ to the line element. But, as will be seen,
this feature gets absorbed in the $e$-basis for $R^{ab}$.)

\item
The second, and more important reason is, that the K-S form is more efficient for
generalisation to axial symmetry~\cite{MP,C1}. In exploring such possibilities, faced
with the complexities of the combinatorics ivolved, it would be helpful to to have ready
the results of the spherically symmetric limit to fall back on as checks.

\end{itemize}

The second of these points concerns however a future project. As will be seen in the
following sections we will also fully exploit the special advantages of \re{13}.

\section{Construction of $p$-Ricci tensors for spherical symmetry and solutions}

For the K-S metric \re{14} the frame vectors are
\begin{eqnarray}
e_{\mu}^a &=& \eta_{\mu}^a +C\: l_{\mu}l^a  \nonumber \\
e^{\mu}_a &=& \eta^{\mu}_a -C\: l^{\mu}l_a \label{a1} \: ,
\end{eqnarray}
and $e^a =e_{\mu}^a \dd x^{\mu}$ with $e_a =\eta_{ab} e^b$.

The spin-connections are
\be
\label{a2}
\omega_{\mu}^{ab}=\eta^{b\nu}\pa_{\nu}e_{\mu}^a -\eta^{a\nu}\pa_{\nu}e_{\mu}^b
\ee
and $\omega^{ab}=\omega_{\mu}^{ab}\: \dd x^{\mu}=-\omega^{ba}$.

Since \re{16} holds for spherical symmetry it is convenient to introduce the notations
\be
\label{a3}
\dd r =\hat x_i \dd x^i \: , \qquad e^r =\hat x_i e^i \: ,
\ee
giving
\begin{eqnarray}
e^0 &=& \dd x^0 -C\: L\: (\dd x^0 +\dd r)  \nonumber \\
e^i &=& \dd x^i +C\: L\: \hat x_i \: (\dd x^0 +\dd r)  \nonumber \\
e^r &=& \dd r + C\: L\: (\dd x^0 +\dd r) \label{a4}
\end{eqnarray}
in which $L=l_o^2$.

Using now \re{4} and \re {a2} one obtains, after straightforward simplifications,
\begin{eqnarray}
R^{0i} &=& A\: e^0 \wedge e^i +B\: \hat x_i \: e^0 \wedge e^r  \nonumber \\
R^{ij} &=& F\: e^i \wedge e^j +H\: e^r \wedge (\hat x_j e^i -\hat x_i e^j) \label{a5}
\end{eqnarray}
where, with $L'=\frac{\dd L}{\dd r}$ and $L''=\frac{\dd^2 L}{\dd r^2}$,
\begin{eqnarray}
A &=& \frac{CL'}{r}\: , \: \qquad B=C(L'' -\frac{L'}{r})=(CL'' -A)  \nonumber \\
F &=& \frac{2CL}{r^2}\: , \qquad H=C(\frac{2L}{r^2}-\frac{L'}{r})=(F-A) \label{a6}.
\end{eqnarray}

Note that the de Sitter solution in $d$ dimensions is obtained by setting
\be
\label{a7}
L=r^2 \: , \qquad A=F=2C\: , \qquad B=H=0 \: ,
\ee
yielding
\be
\label{a8}
R^{ab}=2C\: e^a \wedge e^b \: .
\ee
For higher $p$ members of the hierarchy, one continues to obtain, starting with $p=2$,
\be
\label{a9}
R^{abcd} =3(2C)^2 e^a \wedge e^b \wedge e^c \wedge e^d
\ee
and the situation does not change essentially as $p$ increases, giving at each stage
a constant $R_{(p)}$. Having pointed this out,
we consider henceforth only $p$-flat solutions without cosmological constant.

As a check one may verify, to start with, that for
\[
L=r^{-(d-3)}
\]
one indeed obtains the Schwarzschild solutions in $d$ dimensions~\cite{MP} satisfying
\be
\label{a10}
{R_{(1)}}_b^a \equiv R_b^a =R_{bc}^{ac} =0. 
\ee
This will of course also emerge as the simplest particular ($p=1$) case of our general
solution given below. It is very instructive to study in detail the cases $p=2$ and $3$.
For low values of $d$ one can even write all terms in the summations explicitly. The
general structure is seen to emerge more clearly at each successive stage. For brevity
however, we present directly the general case. From \re{8} and \re{a5} one obtains the
following results.

{\it No summations are involved} in the following particular cases of
\[
R_{b_1 b_2 ...b_{2p}}^{a_1 a_2 ...a_{2p}}
\]
necessary for the passage to ${R_{(p)}}_b^a$. One has

\begin{eqnarray}
R_{i i_1 ...i_{2p-1}}^{0 i_1 ...i_{2p-1}} &=& 0  \label{a11} \\
R_{0 i_1 ...i_{2p-1}}^{i i_1 ...i_{2p-1}} &=& 0  \label{a12} \\
R_{0 i_1 ...i_{2p-1}}^{0 i_1 ...i_{2p-1}} &=& (1.3.5.\: ...(2p-3)) F^{p-2}
\left[ (2p-1)AF + (BF -2(p-1)AH)
(\hat x_{i_1}^2 +...+\hat x_{i_{2p-1}}^2) \right]
 \label{a13} \\
R_{i_1 i_2 ...i_{2p}}^{i_1 i_2 ...i_{2p}} &=& (1.3.5.\: ...(2p-1)) F^{p-1}
\left[ F-H(\hat x_{i_1}^2 +...+\hat x_{i_{2p}}^2) \right]  \label{a14} \\
R_{j_1 i_2 ...i_{2p}}^{i_1 i_2 ...i_{2p}} &=& -(1.3.5.\: ...(2p-1)) F^{p-1}
H\hat x_{i_1}\hat x_{j_1} \: , \qquad i_1 \neq j_1  \label{a15} \\
R_{j_1 0 i_2 ...i_{2p-1}}^{i_1 0 i_2 ...i_{2p-1}} &=& (1.3.5.\: ...(2p-3)) F^{p-2}
\left[ \left(BF-2(p-1)AH\right) \hat x_{i_1}\hat x_{j_1} \right]\: . \label{a16}
\end{eqnarray}

For obtaining ${R_{(p)}}_b^a$ one has to sum over the appropriate indices in each case.
To start with one notes, trivially, from \re{a11} that
\be
\label{a17}
{R_{(p)}}_i^0 =\frac{1}{(2p-1)!} \sum_{i_1 ,...,i_{2p-1}} R_{i i_1 ...i_{2p-1}}^{0
i_1 ...i_{2p-1}} =0\: .
\ee
Similarly, from \re{a12}
\be
\label{a18}
{R_{(p)}}_0^i =0 \: .
\ee
Concerning the other cases it is helpful to note the following points,
\be
\label{a19}
\sum_{i_1 ,...,i_{2p-1}} (\hat x_{i_1}^2 +...+\hat x_{i_{2p-1}}^2 )=\left(
\begin{array}{c}
d-2 \\ 2p-2 \end{array} \right)
\ee
and
\be
\label{a20}
\sum_{i_2 ,...,i_{2p-1}} (\hat x_{i_2}^2 +...+\hat x_{i_{2p-1}}^2 )=\left(
\begin{array}{c}
d-3 \\ 2p-3 \end{array} \right)(1-\hat x_{i_1}^2) \: ,
\ee
where we have used the usual notation
\[
\left( \begin{array}{c}  a \\  b      \end{array}  \right)=
\frac{a!}{b!\: (a-b)!} \: .
\]

For \re{a19} one notes that in summing over the $(2p-1)$-tuples, {\it each} $\hat x_i$
occurs as many times as the possibility of selecting $(2p-2)$-tuples (of distinct
$\hat x_j$'s) among the coordinates after fixing $0$ and $i$, namely among $(d-2)$.
Then one uses
\[
\sum_{i=1}^{d-1} \hat x_i^2 =1 \: .
\]

For \re{a20} one uses analogous arguments, and
\[
\sum_{j \neq i} \hat x_j^2 =1-\hat x_i^2 \: .
\]

We use such results and consider the normalisation factor in \re{11} to be absorbed
implicitly by exhibiting only contributions of distinct numbers of combinations, as in
\re{a19} and \re{a20}. For convenience we also introduce the notations
\begin{eqnarray}
X & \equiv & L^{p-2} \left[ p(rLL')+(d-2p-1)L^2 \right]  \nonumber \\
Y & \equiv & L^{p-2} \left[ r^2 (LL'' +(p-1) L'^2)+(d-2p)(rLL')\right]\: .
\label{a21}
\end{eqnarray}

Using all these results, combining terms and simplifying, completing the necessary
summations and using \re{a5}, one obtains finally,

\begin{eqnarray}
{R_{(p)}}_0^0 &=& (1.3.5...(2p-3))F^{p-2} \left[(2p-1)AF
\left( \begin{array}{c}  d-1 \\  2p-1      \end{array}  \right) +
(BF-2(p-1)AH)\left( \begin{array}{c}  d-2 \\  2p-2      \end{array}  \right)
\right]  \nonumber \\
&=& C^p (1.3.5...(2p-3))\left( \begin{array}{c}  d-2 \\  2p-2      \end{array}  \right)
\frac{2^{p-1}}{r^{2p}}Y  \label{a22}
\end{eqnarray}

\begin{eqnarray}
{R_{(p)}}_j^i &=& (1.3.5...(2p-3)) F^{p-2} \left[ (BF-2(p-1)AH)
\left( \begin{array}{c}  d-3 \\  2p-2      \end{array}  \right)
-(2p-1)FH\left( \begin{array}{c}  d-3 \\  2p-1      \end{array}  \right)
\right] \hat x_i \hat x_j  \nonumber \\
&=& C^p (1.3.5...(2p-3))\left( \begin{array}{c}  d-3 \\  2p-2      \end{array}  \right)
\frac{2^{p-1}}{r^{2p}}(Y-2X)\hat x_i \hat x_j \: ,
\qquad i\neq j   \label{a23}
\end{eqnarray}

\begin{eqnarray}
{R_{(p)}}_i^i &=& (1.3.5...(2p-3))F^{p-2}\: . \nonumber
\\ &.& \: \left[
(2p-1)AF\left( \begin{array}{c}  d-2 \\  2p-2      \end{array}  \right)+ (BF-2(p-1)AH)
\left (\left( \begin{array}{c}  d-2 \\  2p-2      \end{array}  \right)\hat x_i^2 +
\left( \begin{array}{c}  d-3 \\  2p-3      \end{array}  \right)(1-\hat x_i^2) \right)
\right]  \nonumber \\
&+& (1.3.5...(2p-1))F^{p-1}\left[
F\left( \begin{array}{c}  d-2 \\  2p-1      \end{array}  \right)-H\left(
\left( \begin{array}{c}  d-2 \\  2p-1      \end{array}  \right)\hat x_i^2 +
\left( \begin{array}{c}  d-3 \\  2p-2      \end{array}  \right)(1-\hat x_i^2)
\right) \right]  \nonumber \\
&=& C^p (1.3.5...(2p-3)) \frac{(d-3)!}{(2p-2)!(d-2p)!} \frac{2^{p-1}}{r^{2p}}
\left[(p-1)Y+(d-2p)X +(d-2p)(Y-2X)\hat x_i^2 \right] \: .
\label{a24}
\end{eqnarray}

>From \re{a17}, \re{a18}, \re{a22}, \re{a23} and \re{a24} we see that for {\it all} $(a,b)$,
\[
{R_{(p)}}_b^a =0
\]
provided that
\begin{eqnarray}
X &=& L^{p-2} \left[ p(rLL')+(d-2p-1)L^2 \right] =0 \nonumber \\
Y &=& L^{p-2} \left[ r^2 (LL'' + (p-1)L'^2 )+(d-2p)(rLL')\right] =0\: .
\label{a25}
\end{eqnarray}
Since a constant factor of $L$ can be absorbed in $C$,
the {\it single} necessary and sufficient constraint satisfying \re{a25}, and hence the
variational equations \re{a22}-\re{a24}, turns out to be
\be
\label{a26}
L^p =r^{-(d-2p-1)} \: .
\ee
Differentiating \re{a26} once yields
\[
X=0
\]
and differentiating a second time,
\[
Y=0\: .
\]
For $p=1$ this reduces to the standard Schwarzschild metric in $d$ dimensions~\cite{MP} with
\be
\label{a27}
L=l_0^2 =r^{-(d-3)}\: .
\ee

Let us now indicate briefly certain complementary features arising when one uses the diagonal
metric \re{13}. Now one has
\[
\dd s^2 = -N \dd t^2 +N^{-1} \dd r^2 +r^2 \dd \Omega_{(d-2)}^2 \: ,
\]
where
\be
\label{a28}
\dd \Omega_{(d-2)}^2 =\dd \theta_1^2 +\sin^2 \theta_1 \dd \theta_2^2 +...+
(\prod_{n=1}^{d-3}\sin \theta_n)^2 \dd \theta_{d-2}^2\: .
\ee

For a diagonal metric one can set (no summation being involved)
\begin{eqnarray}
e^a &=& \sqrt{g_{aa}} \dd x^a \: ,\qquad x^a =t,r,\theta_1 ,...,\theta_{d-2} \nonumber \\
\omega^{ab} &=& {1\over \sqrt{g_{aa}g_{bb}}}\left[ (\pa_b \sqrt{g_{aa}})e^a
-(\pa_a \sqrt{g_{bb}})e^b \right] \label{a29} \: .
\end{eqnarray}
Thus, in particular,
\begin{eqnarray}
\omega^{\theta_i \theta_j}\equiv \omega^{ij} &=& -
(\cos \theta_i \sin \theta_{i+1} ...\sin \theta_{j-1}) \dd \theta_j \nonumber \\
&=& -{1\over r} \cos \theta_i (\prod_{n=1}^{i}\sin \theta_n)^{-1}e^j =-\omega^{ji} \: ,
\qquad i<j\: . \label{a30}
\end{eqnarray}
These can be shown to satisfy the remarkable relation
\be
\label{a31}
\dd \omega^{ij} +\sum_k \omega^{ik} \wedge \omega^{kj} ={1\over r^2} e^i \wedge e^j \: ,
\qquad 1\le i,j,k \le d-2 \: ,
\ee
independently of $d$.

By virtue of \re{a31} as well as other more evident relations one obtains finally (with index $i$
standing for $\theta_i$)
\begin{eqnarray}
R^{tr} &=& -{1\over 2}N'' \: e^t \wedge e^r \: ,\: \qquad R^{ti}=-{1\over 2r} N'\: e^t \wedge e^i
\nonumber \\
R^{ri} &=& -{1\over 2r}N' \: e^r \wedge e^i \: ,\qquad R^{ij}={1\over r^2}(1-N)\: e^i \wedge e^j
\label{a32}
\end{eqnarray}
with $i,j=1,2,...(d-2)$.

Thus $R^{ab}$ is ``diagonalised'' leading to dramatic simplifications in certain respects. But
as pointed out in section {\bf I}, there are now more distinct blocks to be summed over in
evaluating ${R_{(p)}}_b^a$. Let us illustrate this for the simplest nontrivial case, $p=2$. Now
for ${R_{(2)}}_i^i (\equiv {R_{(2)}}_{\theta_i}^{\theta_i})$ one has {\it four} distinct classes
of terms as compared to {\it two} in the previous case. Thus
\be
\label{a33}
{R_{(2)}}_i^i =\sum_{j}R_{itrj}^{itrj}+\sum_{jk}R_{itjk}^{itjk}+\sum_{jk}R_{irjk}^{irjk}+
\sum_{jkl}R_{ijkl}^{ijkl}\: .
\ee

Implementing \re{a32} in \re{6} one obtains finally, on the right hand side
\be
\label{a39}
\frac{d-3}{2r^4}\left[-r^2 (1-N)N'' +r^2 (-N')^2 -(d-4)r(1-N)N'+(d-4)(1-N)
\left( -2rN'+(d-5)(1-N)  \right)
\right]\: .
\ee
The intermediate, suppressed, steps are straightforward.

Consistently with our previous derivation, this expression vanishes for
\be
\label{a35}
N=N_{(2)}=1-2CL_{(2)}=1-2Cr^{-{1\over 2}(d-5)}\: .
\ee
One can verify that for the general case one indeed obtains
\be
\label{36}
N_{(p)}=1-2CL_{(p)}=1-2Cr^{-\frac{d-2p-1}{p}}\: .
\ee

We will not rederive here this result. But important uses of the diagonal metric are presented
in the following sections.

\section{Maximal extensions and $p$-dependent periodicity for Euclidean signature}

The limiting cases
\[
d=2p \qquad {\rm and} \qquad d=2p+1
\]
will be briefly discussed in the following section. For
\[
d>2p+1
\]
one has a Schwarzschild like event horizon in
\be
\label{b1}
\dd s^2 = -N_{(p)} \dd t^2 +N_{(p)}^{-1} \dd r^2 +r^2 \dd \Omega_{(d-2)}^2 \: ,
\ee
with
\[
N_{(p)}=1-\frac{2C}{r^{\frac{d-2p-1}{p}}} \equiv 1-({K\over r})^{\frac{d-2p-1}{p}}\: ,
\]
at
\[
r=K\: .
\]
One can introduce Kruskal type coordinates, generalised for $p>1$, to desingularise the
horizon. We will be particularly interested in the Euclidean section and the $p$-dependence
of the time-period which becomes necessary for consistency. For the standard case ($d$=4,
$p=1$) this period is well known to be inversely proportional to the Hawking temperature of
the black hole. Here we generalise the treatment (for $d\ge 4$ , $p=1$) presented in
\cite{C2}. For $d=4$ a detailed study citing basic sources is to be found in \cite{BGG}.

We start directly with the Euclidean continuation of \re{b1},
\be
\label{b2}
\dd s^2 = N_{(p)} \dd t^2 +N_{(p)}^{-1} \dd r^2 +r^2 \dd \Omega_{(d-2)}^2 \: .
\ee
Defining
\be
\label{b3}
r^{\star}=\int \frac{\dd r}{N_{(p)}}\: ,
\ee
we introduce the coordinates $(\eta ,\zeta)$ satisfying (for some constant $\lambda$ to be fixed
later)
\be
\label{b4}
\e^{2\lambda r^{\star}} ={1\over 4}(\eta^2 +\zeta^2)\: ,\qquad
\e^{i\lambda t}=\left( \frac{\eta -i\zeta}{\eta +\zeta}\right)^{1\over 2}
\ee
leading to
\be
\label{b5}
\dd s^2 =N_{(p)}(4\lambda^2 \e^{2\lambda r^{\star}})^{-1} (\dd \zeta^2 + \dd \eta^2)
+r^2 \dd \Omega_{(d-2)}^2 \: .
\ee
Setting
\be
\label{b6}
r=K\rho^p \: \: \: \:  {\rm and} \: \: \:\:  n=d-2p-1\: , \qquad r\ge K\: ,\rho \ge 1
\ee
\begin{eqnarray}
r^{\star} &=& \int \frac{\dd r}{1-({K\over r})^{n\over p}}=Kp\int \frac{\rho^{n+p-1}}{\rho^n -1}
\dd \rho \nonumber \\
&=& Kp \left( \int \frac{\dd \rho}{\rho^n -1} +\int \frac{\rho^{n+p-1} -1}{\rho^n -1}
\dd \rho \right)\: .\label{b7}
\end{eqnarray}

It is not necessary for our purposes to evaluate the integrals, though that is possible. As
in \cite{C2} we use
\be
\label{b8}
\frac{1}{x^n -1}={1\over n}\left( \frac{1}{x-1} -\frac{x^{n-2}+2x^{n-3}+...
+(n-2)x+(n-1)}{x^{n-1}+x^{n-2}+...+x+1}\right)
\ee
resulting in
\be
\label{b9}
r^{\star}={Kp\over n}\int\frac{\dd \rho}{\rho -1}+f(\rho)
\ee
where the function $f(\rho)$ which we do not evaluate, does not contribute to the
singularity at $\rho =1$ or $r=K$. Hence
\be
\label{b10}
\e^{-2\lambda r^{\star}}=\left( \left( {r\over K}\right)^{1\over p} -1\right)^{-({2\lambda
Kp\over n})}\e^{-2\lambda h(r)}
\ee
where the function $h(r)$ creates no singularity at the horizon. Choosing
\be
\label{b11}
\lambda =-\left( {n\over 2Kp} \right)=-\left( {d-2p-1\over 2Kp} \right)
\ee
one obtains the required desingularisation at the horizon as a direct generalisation of the
standard case for $p=1$, namely of
\be
\label{b12}
\lambda =-\left( {d-3\over 2K} \right)\: .
\ee

It should be noted that for Lorentz signature there is already an essential singularity at
$r=0$ and for Euclidean signature the domain of real values of $(\eta ,\: \zeta)$ corresponds
to $r\ge K$. {\it Hence in both cases a fractional power of $r$ in $N_{(p)}$ does not
introduce a crucial supplementary complication.} One may also note that though the power of
$r$ in $N_{(p)}$ is in general fractional, for any $p$ and $d$ satisfying
\[
d=(n'+2)p+1\: ,\qquad n'=1,2,3,...
\]
it becomes an integer. Examples of such integers for $p>1$ are
\[
(d\: ,\: p\: ;\: n')=(10\: ,\: 3\: ;1)  ,\: (11\: ,\: 2\: ;\: 3),...
\]

>From \re{b4} and \re{b11} one obtains for $t$ a period
\be
\label{b13}
P_{(p)}={2\pi \over |\lambda|}=\frac{4\pi Kp}{d-2p-1}\: ,\qquad d>2p+1\: .
\ee
For $d=4$, $p=1$, $K=2M$ one gets back to the well-known result for the Schwarzschild metric
\be
\label{b14}
P\equiv P_{(1)}=8\pi M \: .
\ee
For fields on the Euclidean section, periods and temperatures are related inversely through
the Boltzmann constant. Here one sees that, for given $d$, $P_{(p)}$ increases and hence
the temperature decreases as $p$ becomes larger.

\section{Special cases}

We briefly indicate certain features of the special cases $d=2p,\: d=2p+1$, and
$d=4p$.

\subsection{$d=2p$}

It is easily verified that when $d=2p$ the Lagrangian $R_{(p)}$ is a total divergence
and hence can possess topological rather than dynamical properties. In
fact one obtains the Euler number which, if our solution is implemented, is constrained to
be zero -- a new possibility arising for $p> 1$. Here we express the single surviving $2p$-form
in terms of total differentials and note the relation with our general solution.

The simplest example is provided by
\[
d=4\: ,\: \: , p=2\: .
\]
>From \re{a32} and \re{6}
\begin{eqnarray}
R^{tr\theta_1 \theta_2} & =& R^{tr}\wedge R^{\theta_1 \theta_2}+R^{t\theta_2}\wedge
R^{r\theta_1}+R^{t\theta_1}\wedge R^{\theta_2 r} \nonumber \\
&=& {1\over 4r^2}\left[ -(1-N)N''+(N')^2  \right]e^t \wedge e^r \wedge e^{\theta_1}
\wedge e^{\theta_2}\: . \label{c1}
\end{eqnarray}
In conventional notations, $\theta_1 =\theta$, $\theta_2 ={\pi \over 2} -\phi$, and using
$N=1-2CL$,
\be
\label{c2}
R^{tr\theta \phi}=({1\over 2} C^2)\dd t \wedge \frac{\dd}{\dd r} ( \frac{\dd}{\dd r}L^2)
\dd r \wedge \dd \cos \theta \wedge \dd \phi \: .
\ee

For the general case
\[
d=2p\: ,\qquad p=1,2,3,...
\]
the angular factors can again be expressed as total differentials in an evident fashion
while, crucially, the coefficient of $\dd r$ can be shown to become
\be
\label{c3}
\frac{\dd^2}{\dd r^2} (L^p).
\ee
The computation is straightforward but will not be presented here. We just note that this
coefficient vanishes for
\be
\label{c4}
L^p =c_1 +c_2 r
\ee
with $c_1$ and $c_2$ constants. Our generic solution
\be
\label{c5}
L^p =r^{-(d-2p-1)}=r
\ee
for $d=2p$ is included in \re{c4} for $c_1$=0 and $c_2 =1$, giving
\be
\label{c6}
N_{(p)}=1-2Cr^{1\over p}=1-2Cr^{2\over d}\: .
\ee
For $d=4$ this gives
\[
N_{(2)}=1-2Cr^{1\over 2}\: .
\]

In section {\bf III} we generalised Kruskal coordinates to {\it negative} fractional power
of $r$ in $N_{(p)}$. Let us just mention that one can formally generalise Gibbons-Hawking
coordinates \cite{BGG,GH} for such a ``cosmological'' horizon involving a fractional power
${1\over p}$.

Flat space is of course, as always, a solution ($c_1 =c_2 =0$). But for $c_2 =0$ one obtains
a constant $L^p$. This situation is discussed below since it arises generically for
$d=2p+1$.

\subsection{$d=2p+1$}

In this case our generic solution itself gives
\be
\label{c7}
L^p =r^{-(d-2p-1)}=1\: .
\ee
This corresponds to the situation where the $L^2$ term is absent in $X$ given by \re{a21}.

For $2C<1$, now
\be
\label{c8}
N=1-2C=\lambda^2 <1
\ee
and one has still Lorentz signature with
\be
\label{c9}
\dd s^2 =-\lambda^2 \dd t^2 +\lambda^{-2}\dd r^2 +r^2 \dd \Omega_{(d-2)}^2 \: .
\ee
Setting $r=\rho^{\lambda}$ one can obtain an ``isotropic'' form. We do not propose to
discuss further here this degenerate case. Of much more interest is the following one.

\subsection{$d=4p$ : double selfduality}

The general formulation of double selfduality and the generalised Yang-Mills instantons
constructed, for Euclidean signature, in terms of the spin-connections was the major theme
in \cite{OT}. For $d=8$, de Sitter and Fubini-Study metrics were obtained by the present
authors, imposing double selfduality \cite{CT}. Here we will verify that our Schwarzschild
type ($p$-Schwarzschild) solutions also satisfy double selfduality for $d=4p$.

To start with we note that \re{a32} along with \re{8} ensures rhat $R^{a_1 a_2 ...a_{2p}}$
has {\it only one} non-zero component
\[
R_{a_1 a_2 ...a_{2p}}^{a_1 a_2 ...a_{2p}} \: .
\]
This ``diagonalisation'' simplifies dramatically the situation. In the general formulation,
in frame indices, the double self-duality constraint
\be
\label{c10}
R_{b_1 b_2 ...b_{2p}}^{a_1 a_2 ...a_{2p}}=\frac{1}{(2p!)^2}\vep^{a_1 a_2 ...a_{2p}d_1 d_2 ...d_{2p}}
\vep_{b_1 b_2 ...b_{2p}c_1 c_2 ...c_{2p}}R_{d_1 d_2 ...d_{2p}}^{c_1 c_2 ...c_{2p}}
\ee
reduces to
\be
\label{c11}
 R_{a_1 a_2 ...a_{2p}}^{a_1 a_2 ...a_{2p}}
=R_{a_{2p+1} a_{2p+2} ...a_{4p}}^{a_{2p+1} a_{2p+2} ...a_{4p}}
\ee
for {\it all} choices of complementary sets of the $4p$ indices.

For $d=4,\: p=1$
\[
1-N=\frac{2C}{r}
\]
and the well-known double selfduality of the Schwarzschild metric is assured through the
relations
\be
\label{c12}
-{1\over 2}N''={2C\over r^3}={1\over r^2}(1-N)
\ee
since from \re{a32}
\be
\label{c13}
R_{tr}^{tr}=-{1\over 2}N'' \: ,\qquad R_{12}^{12}={1\over r^2}(1-N)\: .
\ee

Even more directly one has
\be
\label{c14}
R_{t2}^{t2}=R_{r1}^{r1}\: ,\qquad R_{t1}^{t1}=R_{r2}^{r2}
\ee
each one being just
\[
-{N' \over 2r}\: .
\]

Thus one has a gravitational instanton on the Euclidean section with nontrivial topological
indices (Euler number and Hirzbruch signature). The spin-connections lead to a selfdual
Yang-Mills field~\cite{B-J}. It is a general fact that such a construction provides a
solution for the combined gravitational-YM system. This is because the metric is not
affected by a {\bf back-reaction} of the YM field, since the stress energy-momentum tensor of
the selfdual YM field vanishes.

For $d=8\: , \: , p=2$, we have
\be
\label{c15}
1-N=2C\: r^{-{1\over 2}(8-4-1)}=2C\: r^{-{3\over 2}}\: .
\ee
This assures in the crucial equality between
\[
R_{tr12}^{tr12}=\left( -{1\over 2}N'' \right)\left( {1\over r^2}(1-N)\right)
+2\left( -\frac{N'}{2r} \right)^2
\]
and
\[
R_{3456}^{3456}=3\left( {1\over r^2}(1-N) \right)^2
\]
and hence the double selfduality constraint
\be
\label{c16}
R_{tr12}^{tr12}=R_{3456}^{3456}\: .
\ee
Here, as before, the indices $1,2,...,6$ refer to $\theta_1 ,\theta_2 ,...,\theta_6$
respectively. Other constraints due to \re{c11} can als be shown to be satisfied
systematically.

For the general case with $d=4p$
\be
\label{c17}
1-N=2C\: r^{-\frac{d-2p-1}{p}}=2C\: r^{-\frac{2p-1}{p}}\: .
\ee
This satisfies the crucial relation
\be
\label{c18}
\left( -{1\over 2}N'' \right)\left( {1\over r^2}(1-N)\right)+(2p-2)\left( -{N' \over 2r}
\right)^2 =(2p-1)\left( {1\over r^2}(1-N)\right)^2 \: ,
\ee
which assures the double selfduality constraint
\be
\label{c19}
R_{tr12...(2p-2)}^{tr12...(2p-2)}=R_{(2p-1)(2p)...(4p-2)}^{(2p-1)(2p)...(4p-2)}\: .
\ee
Other constraints due to \re{c11} can als be shown to be satisfied systematically.

A systematic study of the topological properties of this class of gravitational instantons
and those of the associated YM ones (both for the general $p$ case in the sense of
\cite{OT}) will be presented elsewhere.

\section{A comment on surface deformations and near-horizon symmetry}

A number of recent papers explore ``gauge'' algebra of surface deformations restricted to
event or cosmological horizons~\cite{C,LW}. The references~\cite{C,LW} deal with any
dimension $d$ , but of course for $p=1$. Other important sources, mostly concerning
specific low dimensions, are cited in these papers. From both types of horizons,
the $(r,t)$ plane playing a crucial role, a
Virasoro subalgebra emerges from the study of surface deformations with a central charge
\be
\label{d1}
C=\frac{3A\beta}{2\pi GT}
\ee
where $A$ is the area of the horizon, $(8\pi G)^{-1}$ is the overall factor of the boundary
terms in the full generator of surface deformations, $T$ is the period of rotational
perturbations considered and lastly $\beta$ is obtained as a coefficient on developing the
lapse function $N$ near the horizon. For a Schwarzschid black-hole, for example, $\beta$
turns out to be the inverse of the Hawking temperature~\cite{C}.

For our case, starting with \re{b1}, i.e.
\[
N_{(p)}=1-\left(\frac{K}{r} \right)^{\frac{d-2p-1}{p}}
\]
hence
\[
\frac{\dd N_{(p)}}{\dd r}|_{r=K}=\left( \frac{d-2p-1}{pK}\right)\equiv
\frac{2\pi}{\beta_{(p)}}
\]
when
\be
\label{d2}
\beta_{(p)} =\frac{(2\pi K)\: p}{d-2p-1}\: .
\ee

This is, of course, directly proportional to the period $P_{(p)}$ derived before (see
\re{b13}) and gives for the Schwarzschild case with
\[
d=4\: ,\: \: p=1 \: \: , \: K=2M \: ,
\]
\be
\label{d3}
\beta =4\pi M \; .
\ee

We do not intend to undertake a study of surface deformations for our horizons. We just point
out that, quite plausibly, a $p$-dependent hierarchy of central charges will emerge in our
context.

\section{General remarks}

To construct  our metrics as variational minima of the traces of higher order terms, we have
gone beyond Einsten theory generalised to higher dimensions. Higher order gravitational terms
are nowadays familiar (the so called $(R^2)^2$ and $R^4$ terms) in string theory effective
actions~\cite{GW,K}. But we consider separately the members of one particular
hierarchy of higher order
terms, that in which only velocity-square terms appear, generalising the Einsten-Hilbert
term directly, and not perturbative contributions, starting with the standard formalism. In
certain constructions of topological field theories~\cite{B} higher order terms are
introduced directly at the start. One may keep such aspects in mind. But the real motivation
for presenting our formal structure is the discovery of  the remarkable class of solutions
with their simple, suggestive and beautiful properties discussed above. Thus, for example,
rich topological possibilities involving generalised gravitational and YM instantons enter
in the wake of our solutions merely as a particular case ($d=4p$). More generally, topological
aspects in higher dimensions~\cite{B,EGH} deserve study specifically in the context of our
formulation.

An adequate formulation of surface deformation algebras, alluded to in Sec. {\bf V}, might
be of interest. We intend to explore elsewhere possibilities of axially symmetric, stationary
solutions in our context. All these aspects along with a deeper understanding  of the role
of higher order forms provide an interesting program.

\bigskip
\bigskip

\noindent
{\bf \large Acknowledgements}

One of us (A.C.) acknowledges with thanks, discussions with
J.P. Bourguignon, S.F. Hassan, J. Lascoux and
B. Pioline. This work was carried out in the framework of the Enterprise-Ireland/CNRS
programme, under project FR/99/025.

\bigskip
\bigskip

\small{

 }

\end{document}